\documentstyle[prl,twocolumn,aps]{revtex}
\begin{document}
\noindent
{\bf Hellberg and Manousakis reply:}
In our Letter \cite{hellberg99stripe} we concluded
that the ground state of the {\em t-J} model does
not have stripes at $J/t=0.35$.
In the preceding Comment \cite{white00}, White and Scalapino (WS)
raise several objections to our findings.
We refute all of their points and argue that our analysis is indeed correct.

The physical mechanism cited by 
WS explains why
{\it if a striped state were the ground state of the {\em t-J} model}, 
the domain boundaries
would prefer a $\pi$-phase shift in the antiferromagnetic background.
However, this 
mechanism does not favor the stripe state over other
alternative candidates for the ground state of 
the {\em t-J} model in the low hole-doping region. One
serious candidate is a state of electronic phase 
separation\cite{emery90,hellberg97} where an antiferromagnetic
region is separated from the hole rich region by only one 
(as opposed to infinitely many as is the case for the
striped state) energetically costly interface. 
When a {\it finite-size} system is
in a region of the parameter space for which the infinite system would
phase separate, its energy
will be best minimized if the two components in which
the finite-system tends to phase-separate respect the
geometry imposed by the boundaries. 
Thus an instability or near instability to phase separate
may cause domain walls or other structures to form in a {\it finite-size} system.
In
the thermodynamic limit,
the strong fluctuations in the one-dimensional stripes
will destroy the stripes.
Such fluctuation effects are suppressed in finite-size systems where 
only a few stripes are present and the length of the stripes is very limited.
The argument cited by WS 
explains why if one has stripes 
one has to have a $\pi$-phase shift in the antiferromagnetic
order parameter to accommodate hole motion.
It does not explain why stripes are formed 
as opposed to a phase separated state. 

WS 
view their boundary condition as 
a symmetry breaking field whose strength can be taken to
zero. However, this procedure requires taking the
thermodynamic limit and studying whether or not
the stripes remain. 
By studying WS's results obtained for cylindrical boundary conditions 
as a function of the 
number of legs in the cylinder,
it seems that the stripes are strongly influenced
by finite size effects.
Depending on the cylinder's width,
WS find stripes with different linear hole densities along the stripe.
In six-leg ladders, 
the optimum linear density is $\rho_6 = 2/3$ \cite{white99a},
and in eight-leg ladders, $\rho_8 = 1/2$ \cite{white98}.

WS find that the addition of a $t'$ term destroys stripes 
in their calculations.
A next-nearest-neighbor hopping $t'$ inhibits 
phase separation in the {\em t-J} model.
If in a particular finite geometry, 
the near instability to phase separate is manifested 
by stripe formation, adding a $t'$ hopping 
will destabilize the stripes.

WS are incorrect in stating that our conclusions
would have been different if we had excluded the clusters
which they believe are too one dimensional.
The (2,2) translation vectors are four lattice spacings in
distance, just as the (0,4) translation vectors are.
However, irrespective of which clusters we keep or exclude, 
our conclusions are unchanged.  
Even if we restrict ourselves to cluster No.\ 2, 
the lowest energy state of this cluster has energy 
${\cal E}_0 = -0.6397t$ and is not striped.
State (e), the lowest energy vertical stripe state, has energy 
${\cal E}_e = -0.6353t$.
Thus even if we had only examined the cluster in which we found
the vertical stripes, we would still conclude that these stripes
are excited states. 

WS believe a system with at least four holes is required
to study stripe formation.
If many-hole correlations are important, one needs to 
explain why stripes are seen in mean-field studies.
Prelovsek and Zotos \cite{prelovsek93} only found stripe correlations
for large $J/t$ and did not study stripe correlations in two-hole 
clusters.
The degenerate excited states (e) and (f) in our periodic cluster No.\ 2
are nearly identical to the stripes seen by WS \cite{white98}.
The charge density wave amplitude and the spin structure (including
the $\pi$-phase shift) are the same.
And, as shown in Fig.\ 3 of our Letter \cite{hellberg99stripe},
the use of open boundary conditions in one direction breaks the degeneracy and
stabilizes these states as the ground state.

WS argue that our results might be a finite-size effect (FSE);
however, our main reason for performing a small cluster exact calculation
was to 
study the role of FSEs on the formation of stripes.
Thus, FSEs are {\it welcome} in our calculation.
Since stripes are periodic structures, calculations on 
small periodic clusters that are commensurate with the 
stripe order including the $\pi$-phase shift between 
stripes, such as our cluster No.\ 2, {\em favor} the 
formation of stripes.
We do find stripes in cluster No.\ 2 with exactly the same structure as those 
found by WS but only as excited states.
The fact that these stripes are not the ground state even 
of the cluster most favorable for their formation indicates 
that the ground state of the infinite system is not striped.

We thank S.A. Kivelson for useful discussions.
This work was supported by the Office of Naval Research.

~

\noindent C. Stephen Hellberg$^1$ and E. Manousakis$^2$

$^1$Center for Computational Materials Science

~ Naval Research Laboratory, Washington, DC 20375

$^2$Department of Physics and MARTECH

~ Florida State University,  Tallahassee, FL 32306-3016

\vspace{0.1in}
\today

\end{document}